\documentclass[conference]{IEEEtran}
\IEEEoverridecommandlockouts
\usepackage{cite}
\usepackage{amsmath,amssymb,amsfonts}
\usepackage{algorithmic}
\usepackage{graphicx}
\usepackage{textcomp}
\usepackage{xcolor}
\usepackage[frozencache=true,cachedir=minted]{minted}
\usepackage{tikz}
\usetikzlibrary{shapes.multipart, arrows.meta}
\usepackage{hyperref}
\def\BibTeX{{\rm B\kern-.05em{\sc i\kern-.025em b}\kern-.08em
    T\kern-.1667em\lower.7ex\hbox{E}\kern-.125emX}}
\begin{document}

\title{Observing the Quantum Compiler through Automatic Experiment Tracking for Qiskit\\
\thanks{This work has been supported by the Academy of Finland (project DEQSE 349945) and Business Finland (EM4QS 155/31/2024).}
}

\author{\IEEEauthorblockN{Vlad Stirbu}
\IEEEauthorblockA{\textit{University of Jyväskylä} \\
Jyväskylä, Finland \\
vlad.a.stirbu@jyu.fi}
\and
\IEEEauthorblockN{Arianne Meijer–van de Griend}
\IEEEauthorblockA{\textit{University of Helsinki} \\
Helsinki, Finland \\
ariannemeijer@gmail.com}
}

\maketitle

\begin{abstract}
Understanding the effectiveness of quantum compilation techniques requires visibility into the entire transpilation process, not just the final circuit metrics. This demonstration presents an MLflow-inspired autologging framework for Qiskit that automatically captures compiler provenance, including transpilation stages, pass-level execution data, backend characteristics, compiler configuration, and execution results. The framework extends the QProv provenance model with compiler-specific information and stores the collected data in an MLflow Tracking Server for analysis and visualization. By eliminating manual instrumentation, the proposed approach improves compiler observability and supports reproducible evaluation of quantum compilation workflows.
\end{abstract}


\section{Introduction}

Quantum compilation has become a critical component of the quantum software stack, bridging high-level quantum programs and the hardware-specific circuits that can be executed on today's quantum processors. As quantum compilers continue to evolve, researchers increasingly evaluate new synthesis, mapping, routing, and optimization techniques targeting diverse hardware architectures and compiler configurations. Consequently, benchmarking and resource estimation have become central activities in quantum compilation research, requiring software tools that support systematic experimentation and reproducible evaluation. 

Despite this need, evaluating compilation techniques requires considerably more than reporting the final circuit depth, gate count, or execution fidelity. Understanding why a compiler produced a particular result requires visibility into the complete compilation process: which transpiler passes were executed, how each pass transformed the circuit, the execution time of individual passes, the characteristics of the target backend, the compiler configuration, and the provenance of the entire experiment. Although frameworks such as Qiskit expose much of this information internally, collecting it currently requires substantial manual instrumentation, making comprehensive experiment tracking difficult to adopt consistently and increasing the effort required to reproduce and compare compilation studies.

This paper presents an automatic compiler observability tool for Qiskit that brings MLflow-inspired autologging capabilities to quantum compilation experiments. By transparently instrumenting the transpilation workflow, the tool automatically records compiler provenance without requiring extended modifications to the user code. The collected information is integrated into a structured experiment log and complemented by visualizations, such as transpilation timelines, that help researchers analyze compiler behavior, compare optimization strategies, and reproduce quantum compilation experiments. %

\section{Design}

\subsection{Principles}

The design of the proposed framework is guided by three principles that aim to balance ease of adoption with comprehensive compiler observability.

\begin{itemize}
    \item \textbf{Minimal developer effort}. The primary objective is to minimize the effort required to instrument quantum applications. Inspired by MLflow autologging, the framework automatically intercepts key stages of the Qiskit transpilation and execution workflow, collecting provenance transparently without requiring developers to modify their application code. Enabling experiment tracking therefore consists of a single call to activate autologging, allowing existing quantum programs to benefit from comprehensive provenance collection with minimal changes.

    \item \textbf{Comprehensive provenance}. The framework is designed to capture the complete context of a compilation experiment while remaining compatible with the QProv provenance model. Existing QProv entities describing the quantum program, execution environment, backend, and results are preserved, while compiler-specific information is incorporated through schema extensions rather than modifications. In particular, the framework augments the Compilation section with transpilation stage plans and pass-level execution records, providing sufficient information to reconstruct and analyze the compilation process while maintaining interoperability with QProv-based tools and analyses.

    \item \textbf{Extensibility to other SDKs}. Although the current implementation targets Qiskit, the architecture is designed to support additional quantum software development kits. The autologging functionality is implemented independently of application logic, allowing SDK-specific instrumentation to be added while preserving a common provenance model and experiment storage format. This separation facilitates future support for platforms such as Qrisp\footnote{\href{https://www.qrisp.eu}{https://www.qrisp.eu}} and PennyLane\footnote{\href{https://pennylane.ai}{https://pennylane.ai}}, enabling a consistent observability experiment tracking experience across different quantum programming ecosystems.
\end{itemize}

\begin{figure}
    \centering
    \includegraphics[width=0.7\linewidth]{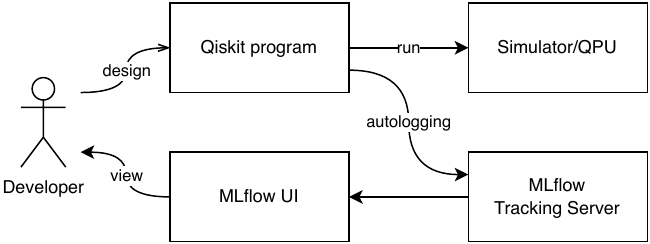}
    \caption{System architecture}
    \label{fig:arch}
\end{figure}

\subsection{System architecture}

MLflow \cite{zaharia2018accelerating} provides a mature experiment tracking infrastructure including structured metadata storage, artifact management, comparison of multiple runs, and an interactive user interface. Reusing this infrastructure avoids developing a dedicated provenance database while enabling immediate analysis capabilities familiar from machine learning. Although other experiment management platforms\footnote{\href{https://github.com/AqueductHub}{https://github.com/AqueductHub}} have also been explored for quantum computing, they provide a less mature ecosystem for provenance tracking and lack the broad community adoption and extensibility of MLflow for integrating compiler observability into existing quantum software workflows.

The proposed system architecture consists of four main components: (1) a Python-based quantum application developed using Qiskit, (2) an automatic observability layer that instruments the application through MLflow-inspired autologging, (3) an MLflow Tracking Server for persisting experiment metadata and provenance information, and (4) the MLflow user interface for exploring and analyzing the recorded experiments. The autologging layer transparently intercepts key stages of the Qiskit transpilation and execution APIs, automatically collecting compiler provenance and execution results as the application is transpiled and executed on either a simulator or a physical quantum processing unit (QPU). The collected information is stored as an MLflow experiment, enabling reproducible execution, systematic comparison of compilation runs, and interactive visualization of compiler behavior without requiring manual instrumentation of the application code. The system architecture is depicted in Fig.~\ref{fig:arch}.

\subsection{Logged information}

\newcommand{\new}[1]{\textcolor{green!50!black}{#1}}

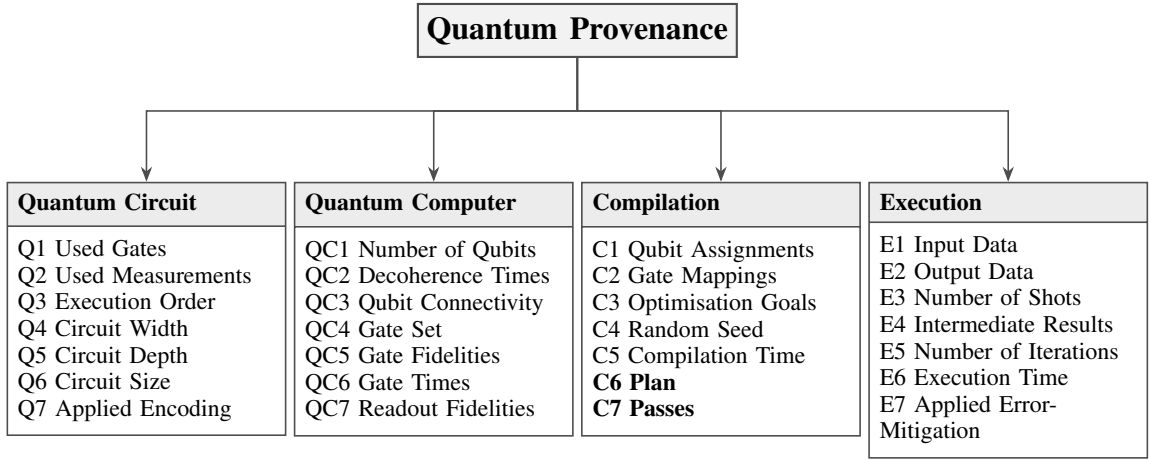
\begin{figure*}[t]
\centering
{\fontfamily{ptm}\selectfont
\begin{tikzpicture}[
    box/.style={
        rectangle split,
        rectangle split parts=2,
        draw=black!70,
        text width=3.4cm, %
        align=left,
        font=\small,
        rectangle split part fill={gray!15,white},
        rectangle split part align={center,left},
        inner sep=4pt,
        anchor=north,
        line width=0.6pt,
        minimum height=4.8cm
    },
    arrow/.style={-Stealth, semithick, black!70}
]

\node[draw=black!70, fill=gray!10, line width=0.8pt, minimum width=3.4cm, minimum height=0.7cm] 
    (root) at (0,1.5) {\large\textbf{Quantum Provenance}};

\node[box] (qc) at (-5.7,-0.5) { %
    \textbf{Quantum Circuit}
    \nodepart{two}
    Q1 Used Gates\strut \\
    Q2 Used Measurements\strut \\
    Q3 Execution Order\strut \\
    Q4 Circuit Width\strut \\
    Q5 Circuit Depth\strut \\
    Q6 Circuit Size\strut \\
    Q7 Applied Encoding\strut
};

\node[box] (qcomp) at (-1.9,-0.5) { %
    \textbf{Quantum Computer}
    \nodepart{two}
    QC1 Number of Qubits\strut \\
    QC2 Decoherence Times\strut \\
    QC3 Qubit Connectivity\strut \\
    QC4 Gate Set\strut \\
    QC5 Gate Fidelities\strut \\
    QC6 Gate Times\strut \\
    QC7 Readout Fidelities\strut
};

\node[box] (comp) at (1.9,-0.5) { %
    \textbf{Compilation}
    \nodepart{two}
    C1 Qubit Assignments\strut \\
    C2 Gate Mappings\strut \\
    C3 Optimisation Goals\strut \\
    C4 Random Seed\strut \\
    C5 Compilation Time\strut \\
    \textbf{C6 Plan}\strut \\
    \textbf{C7 Passes}\strut
};

\node[box] (exec) at (5.7,-0.5) { %
    \textbf{Execution}
    \nodepart{two}
    E1 Input Data\strut \\
    E2 Output Data\strut \\
    E3 Number of Shots\strut \\
    E4 Intermediate Results\strut \\
    E5 Number of Iterations\strut \\
    E6 Execution Time\strut \\
    E7 Applied Error-Mitigation\strut
};

\draw[arrow] (root.south) -- ++(0,-0.7) -| (qc.north);
\draw[arrow] (root.south) -- ++(0,-0.7) -| (qcomp.north);
\draw[arrow] (root.south) -- ++(0,-0.7) -| (comp.north);
\draw[arrow] (root.south) -- ++(0,-0.7) -| (exec.north);

\end{tikzpicture}
}
\caption{Quantum provenance taxonomy capturing key metadata across the quantum computing stack, adapted from \cite{weder2021qprov}. Items marked bold represent extensions added to support the compiler observability case.}
\label{fig:quantum_provenance}
\end{figure*}

The information captured by the proposed tool builds upon the provenance model defined by QProv \cite{weder2021qprov}, ensuring that the recorded experiments are compatible with an existing quantum provenance schema while extending it with compiler-specific information. The base set of logged attributes includes metadata describing the quantum program, execution environment, backend, transpilation configuration, execution results, and generated artifacts. This provides a comprehensive description of the experiment that supports reproducibility and comparison across compilation runs.

To improve compiler observability, we extend the Compilation section of QProv with two additional entities: C6 Plan and C7 Passes. The \textbf{C6 Plan} records the transpilation plan constructed by the compiler prior to execution. It captures the sequence of compilation stages together with the associated compiler passes and any control-flow constructs (e.g., conditional execution or iterative optimization loops) that determine how the transpiler processes the quantum circuit. This information exposes the intended compilation strategy independently of the actual runtime behavior.

The \textbf{C7 Passes} entity captures the outcome of each compiler pass executed during transpilation. For every pass, the tool records the resulting circuit together with key circuit metrics inherited from QProv (Q4–Q7 and C5), as well as compiler-specific metadata including the pass name and the transpilation stage to which the pass belongs. Collectively, these records enable detailed analysis of how individual passes transform the circuit, identify performance bottlenecks, and reconstruct the complete compilation process for visualization and debugging purposes.

\section{Demonstration}

\subsection{How to use}

\begin{figure}
    \centering
    \begin{minted}[fontsize=\scriptsize, breaklines=true, highlightlines={3, 5,10,13}]{python}
import mlflow
from mqt.bench import BenchmarkLevel, get_benchmark
from q8s.runtime.mlflow.qiskit import autolog

autolog()

from iqm.qiskit_iqm.fake_backends.fake_aphrodite import IQMFakeAphrodite
from qiskit.transpiler import generate_preset_pass_manager

mlflow.set_experiment("qiskit-transpilation")


with mlflow.start_run():
    qc = get_benchmark(
        benchmark="qft",
        level=BenchmarkLevel.ALG,
        circuit_size=30,
    )

    backend = IQMFakeAphrodite()

    manager = generate_preset_pass_manager(
        optimization_level=3, backend=backend, seed_transpiler=42
    )

    tqc = manager.run(qc)

    job = backend.run(tqc, shots=1024, memory=True)

    result = job.result()

    result.get_counts()

    \end{minted}
    \caption{Sample Qiskit program with compiler observability. Highlighted lines show changes needed to enable the functionality.}
    \label{fig:qiskit-code}
\end{figure}

Using the proposed compiler observability framework requires only enabling the autologging functionality before executing a standard Qiskit workflow. Fig.~\ref{fig:qiskit-code} shows a minimal example in which a quantum circuit generated with the Munich Quantum Toolkit (MQT) Bench \cite{mqt} is transpiled and executed on an IQM simulator. Once autologging is enabled, the framework transparently instruments the transpilation and execution pipeline, requiring no modifications to the application logic.

During execution, the MLflow experiment run and records the information defined by QProv together with the compiler observability extensions introduced in this paper. The collected provenance is stored in the MLflow Tracking Server, where it can be inspected through the web interface or consumed programmatically for further analysis.

\begin{figure*}
    \centering
    \includegraphics[width=\linewidth]{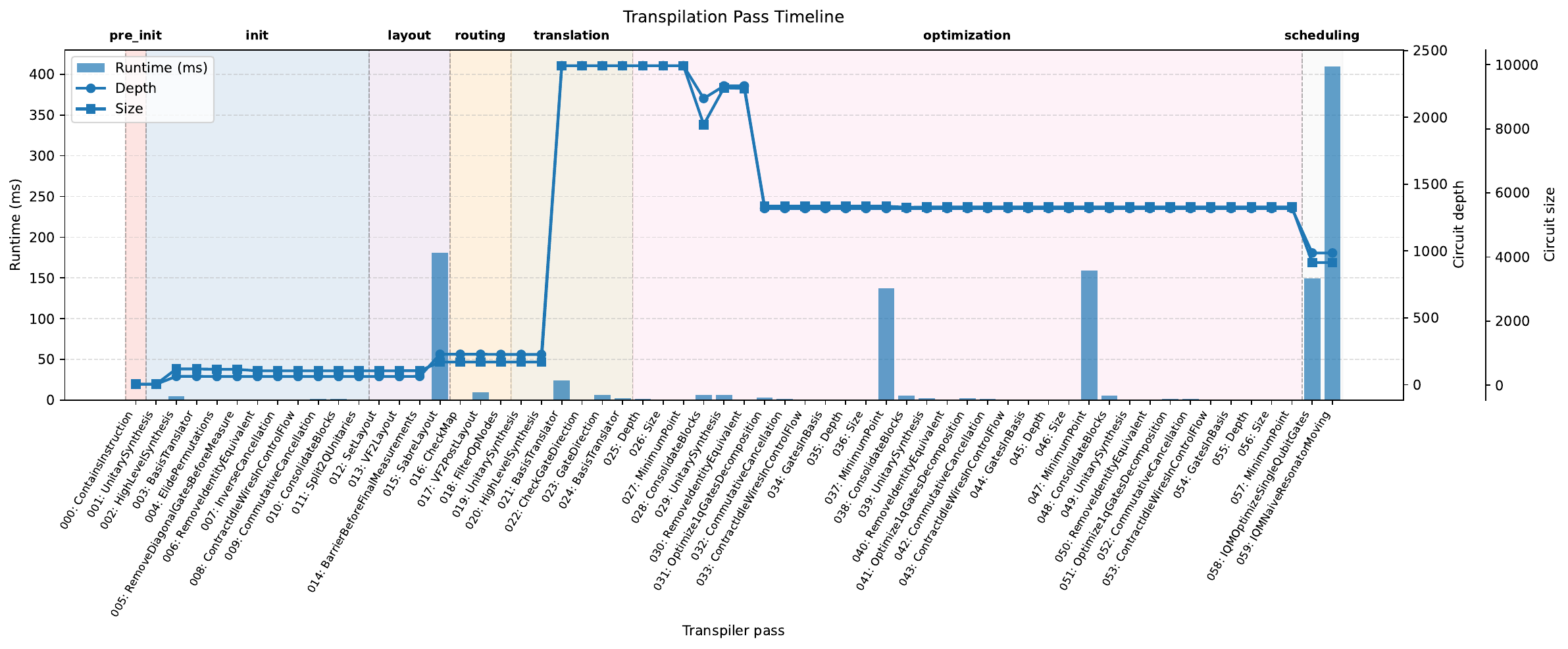}
    \caption{Transpilation pass timeline rendering based on the information collected by Qiskit autologger}
    \label{fig:qiskit-preset-timeline}
\end{figure*}

This workflow preserves the familiar Qiskit programming model while eliminating the manual instrumentation traditionally required to capture detailed compiler provenance. Consequently, researchers can focus on evaluating compilation techniques rather than implementing custom logging infrastructure, making it easier to reproduce experiments and compare different compiler configurations.

\subsection{Visualizations}

The collected provenance data and derived visualization is collected as artifacts and can be explored through the MLflow user interface or rendered using dedicated components. Fig.~\ref{fig:qiskit-preset-timeline} presents a transpilation timeline that combines pass execution times, circuit evolution, and compiler stages, providing an intuitive view of how the transpiler transforms the circuit throughout the compilation process. Such visualizations help identify expensive compiler passes, understand the impact of individual transformations, and compare different transpilation strategies. Beyond the timeline presented in this demonstration, the recorded provenance can be used to generate a variety of other visualizations, including compiler pipeline graphs, pass dependency graphs, circuit metric evolution, backend performance comparisons, and dashboards for analyzing multiple compilation experiments.

\subsection{Experiment comparisons with MLflow}

The MLflow user interface enables researchers to compare multiple compilation experiments without requiring custom analysis tools. Since each transpilation run is recorded as a separate experiment, the UI provides tabular and graphical views of the captured parameters, metrics, and artifacts, allowing differences between compiler configurations to be identified quickly.

\begin{figure}
    \centering
    \includegraphics[width=\linewidth]{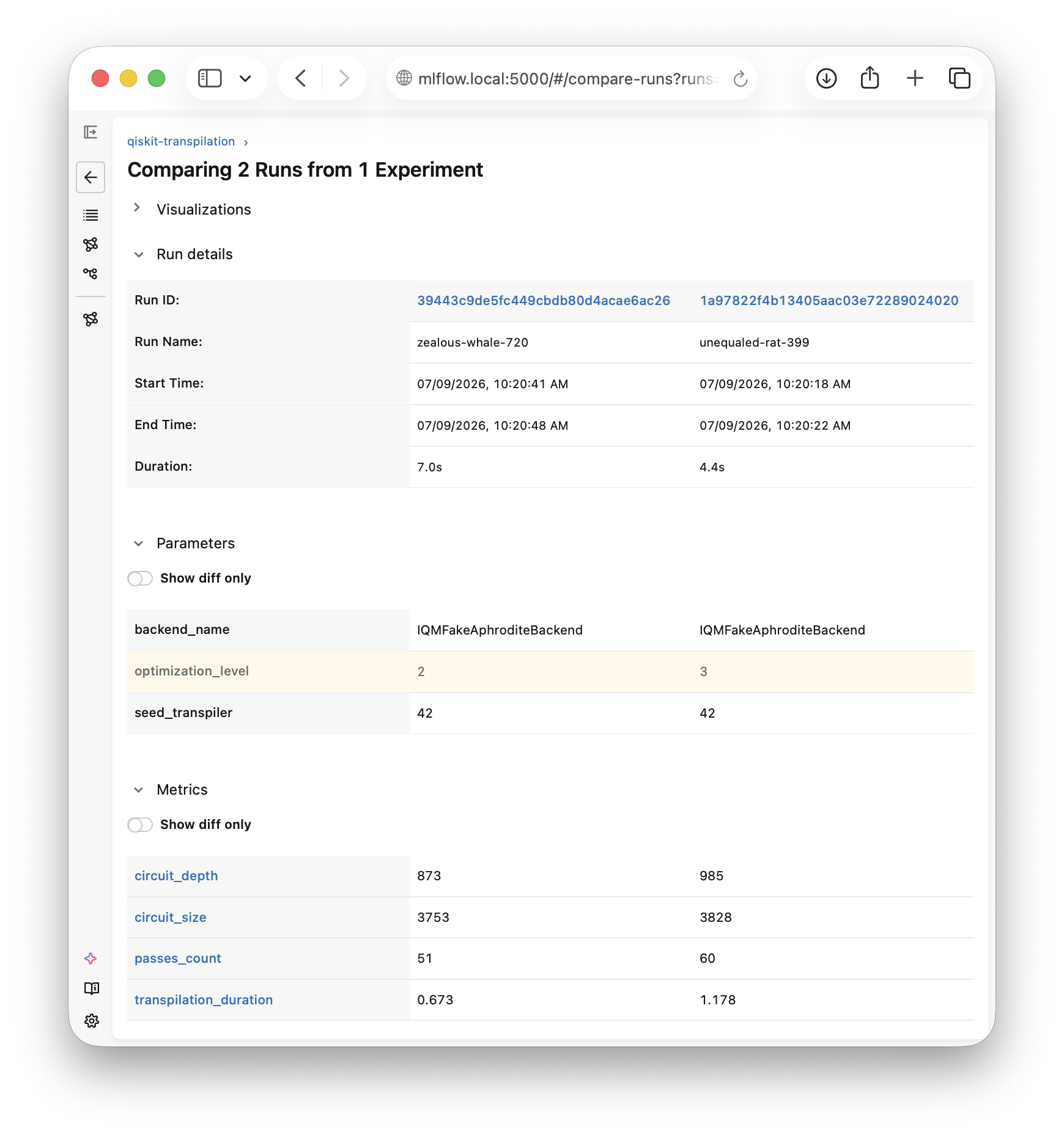}
    \caption{Visualizing experiment comparison metrics with MLflow UI}
    \label{fig:mlflow-ui}
\end{figure}

For example, Fig.~\ref{fig:mlflow-ui} illustrates this capability by comparing two executions of the same quantum circuit transpiled using optimization levels 2 and 3. While conventional experiment tracking reveals that optimization level 3 produces a smaller circuit than level 2, it does not explain why. By selecting either run, the collected pass-level provenance and transpilation timeline expose which compiler passes were executed, where circuit depth changed, and which optimization stages dominated compilation time. This enables researchers to attribute improvements to specific compiler transformations rather than only comparing final metrics.

\section{Contributions}

This demonstration paper presents an automatic compiler observability framework that improves the observability of the quantum compilation process. The work makes the following contributions:

\begin{itemize}
    \item \textbf{Automatic compiler observability for Qiskit}, through MLflow-inspired autologging enabling transparent collection of provenance information from transpilation and execution workflows in an unified experiment record without requiring manual instrumentation of user applications.
    \item \textbf{Extensions to the QProv provenance model} that capture compiler-specific information, including the transpilation stage plan and pass-level execution records, allowing the complete compilation process to be reconstructed.
    \item \textbf{Visualization support} that transforms the collected provenance into interactive views, such as transpilation timelines, demonstrating how experiment tracking can aid in understanding, debugging, and comparing compilation strategies.
\end{itemize}

Together, these contributions provide a practical foundation for reproducible quantum compilation experiments while reducing the effort required to instrument and analyze Qiskit-based workflows.

\section{Tool availability}

The framework is publicly available as open-source software. The source code is hosted in a Git repository\footnote{\href{https://github.com/qubernetes-dev/runtime}{https://github.com/qubernetes-dev/runtime}}, while the package is distributed through the Python Package Index (PyPI), allowing installation via pip\footnote{\href{https://pypi.org/project/q8s.runtime/}{https://pypi.org/project/q8s.runtime/}}.

\section{Conclusions and future work}

This paper presented an observability framework for quantum compilation that brings MLflow-inspired autologging to Qiskit workflows. By automatically capturing compiler provenance, the proposed approach improves the observability and reproducibility of quantum compilation experiments while eliminating the need for manual instrumentation. As this work is intended as a community tool, we are particularly interested in feedback from quantum software practitioners regarding the usefulness of the collected provenance, the proposed visualizations, and additional compiler insights that would support their development and research workflows. We hope this work encourages broader adoption of reproducible, observable quantum compiler experimentation.

Future work will focus on extending the autologging framework beyond Qiskit to support additional quantum software development kits, including Qrisp and PennyLane. This will enable a unified experiment tracking experience across multiple quantum programming ecosystems and facilitate comparative studies of compilation and execution workflows. We also plan to investigate additional visualization techniques and analyses that further enhance compiler observability and support the debugging and optimization of quantum software.


\end{document}